\documentclass[aps,preprint,amssymb,superscriptaddress]{revtex4}

\usepackage{graphicx}
\usepackage{dcolumn}
\usepackage{bm}
\usepackage{color}

\begin{document}


\title{A high-frequency electron paramagnetic resonance spectrometer
for multi-dimensional, -frequency and -phase pulsed measurements}

\author{F. H. Cho}
\affiliation{Department of Physics, University of Southern California Los Angeles CA 90089}%
\author{V. Stepanov}
\affiliation{Department of Chemistry, University of Southern California Los Angeles CA 90089}%
\author{S. Takahashi}
\email{susumu.takahashi@usc.edu}
\affiliation{Department of Chemistry, University of Southern California Los Angeles CA 90089}%
\affiliation{Department of Physics, University of Southern California Los Angeles CA 90089}%

\date{\today}

\begin{abstract}
We describe instrumentation for a high-frequency electron paramagnetic resonance (EPR) and pulsed electron-electron double resonance (PELDOR) spectroscopy. The instrumentation is operated in the frequency range of 107$-$120 GHz and 215$-$240 GHz and in the magnetic field range of 0$-$12.1 Tesla.
The spectrometer consisting of a high-frequency high-power solid-state source, a quasioptical system, a phase-sensitive detection system, a cryogenic-free superconducting magnet and a $^4$He cryostat enables multi-frequency continuous-wave EPR spectroscopy as well as pulsed EPR measurements with a few hundred nanosecond pulses.
Here we discuss the details of the design and the pulsed EPR sensitivity of the instrumentation.
We also present performance of the instrumentation in unique experiments including PELDOR spectroscopy to probe correlations in an insulating electronic spin system and application of dynamical decoupling techniques to extend spin coherence of electron spins in an insulating solid-state system.
\end{abstract}

\maketitle

\section{INTRODUCTION}
Electron paramagnetic resonance (EPR) or electron spin resonance (ESR) spectroscopy has been widely applied
to probe and study local structures and dynamic properties of various compounds in liquids and solids;
for example, structures and dynamics in biological molecules~\cite{Hubbell96, Hubbell00, ESRbio07}, magnetic structures and relaxations in magnetic molecules~\cite{hill98, Barra00, Bertaina08, Takahashi11} and quantum coherence in solid-state spin systems~\cite{Gaebel06, Takahashi08, Hanson07, Lyon06}.
Similar to nuclear magnetic resonance (NMR) spectroscopy,
EPR spectroscopy at high frequencies is highly advantageous to improve the spectral resolution, spin polarization, sensitivity and time resolution of pulsed EPR~\cite{Freed00}. In addition, it has recently been demonstrated that high-frequency EPR (HFEPR) spectroscopy increases spin coherence with decreasing temperature in various spin systems~\cite{Takahashi08, Takahashi09prl, Vantol09, Edwards12}.

Several pulsed HFEPR spectrometers above 150 GHz has been constructed around the world, including Frankfurt~\cite{Hertel05}, Berlin~\cite{Fuhs02}, Leiden~\cite{Blok08} and Tallahassee~\cite{Vantol05, Morley08}.
In addition, a 263 GHz EPR spectrometer has recently been commercially available from Bruker Biospin~\cite{Bruker263GHz}.
A trend of pulsed HFEPR spectroscopy is toward enhancement of excitation power to improve the time resolution and applications of multi-frequency, multi-dimensional and sophisticated pulsed techniques~\cite{Cruickshank09, Freed00, Pannier00, Saxena96, Borbat99, Morley08}. A high-power HFEPR spectrometer based on a free-electron laser has been developed recently to realize pulsed EPR measurements with several nanosecond pulses~\cite{Takahashi07, Takahashi09, Takahashi12}.

In this article, we present the development of a pulsed HFEPR spectrometer powered by a high-frequency high-power solid-state source.
The frequency of solid-state source is tunable in 107-120 GHz and 215-240 GHz.
The solid-state source also enables multiple pulsed excitations with a phase control,
which make it possible to apply more sophisticated pulsed EPR techniques such as dynamical decoupling (DD).
It is also capable of applying multi-dimensional and -frequency pulsed techniques such as pulsed electron-electron double resonance (PELDOR) spectroscopy.
The HFEPR spectrometer employs a quasioptical system for efficient transmission of excitations and EPR signals as well as high isolation of EPR signals from the excitations.
EPR signals are detected by a phase-sensitive superheterodyne detection system.
We discuss the sensitivity of pulsed EPR measurements by considering the single-to-noise ratio of spin echo signals.
In addition, we show unique applications of the HFEPR spectrometer including PELDOR spectroscopy on nitrogen paramagnetic impurities in diamond to detect coupling between nitrogen spins and DD techniques to extend spin coherence of nitrogen impurities from 23 to 211 $\mu$s.

\section{DESIGN AND IMPLEMENTATION}
An overview of the HFEPR spectrometer at USC is shown in Fig~1.
Outputs of a high-frequency high-power (the peak power of 100/700 mW at 230/115 GHz respectively) solid-state source for continuous-wave (cw) and pulsed excitations are first propagated in a quasioptical transmitter stage consisting of an isolator, a circulator and quasioptical mirrors,
then propagated through a corrugated waveguide to couple to a sample located at the center of 12.1 Tesla cryogenic-free superconducting magnet.
The spectrometer employs the induction mode detection for EPR detection ~\cite{smith98} where the cross-polarized component of EPR signals
is separated from the input using a circulator based on a wire-grid polarizer.
EPR signals are then detected by a mixer detector in a quasioptical receiver stage.
Detected EPR signals are down-converted by mixing with a local oscillator (LO) in the receiver stage to produce intermediate frequency (IF) signals at 3 GHz.
In the detection system, IF signals are further mixed with a 3 GHz reference synthesized from the source and the LO to obtain d.c. in-phase (I) and quadrature (Q) signals.
In pulsed EPR measurements, I and Q components of transient EPR signals are sampled by a 2 GSamples/s digital oscilloscope (Agilent).
In cw EPR measurements, intensities of I and Q signals at a field-modulation frequency (20$-$100 kHz) are measured using lock-in amplifiers (Stanford Research Systems).
A pulse generator (SpinCore), magnet power supplies, data acquisition from the digital oscilloscope and lock-in amplifiers are controlled by National Instruments LabVIEW VIs in a control computer.

\subsection{High-frequency high-power solid-state source}
Figure~2 shows a circuit diagram of the high-frequency high-power solid-state source custom-built by Virginia Diodes, Inc (VDI).
The output frequency of the source is tunable in between 215$-$240 and 107$-$120 GHz.
The high-frequency source consists of two microwave synthesizers (8$-$10 GHz and 9$-$11 GHz, Micro Lambda Wireless), isolators (Ditom Microwave), fast p-i-n (PIN) switches (American Microwave), directional couplers (ATM), power splitters and combiners (Narda Microwave East), variable phase shifters (ARRA), preamplifiers and frequency multipliers.
As shown in Fig.~2(a), the single-frequency configuration, {\it e.g.} for cw EPR, spin echo and DD measurements, utilizes one synthesizer.
Output of the synthesizer is first connected to the directional coupler to provide a reference for the detection system (Reference 1 in Fig.~2), then to the isolator.
After the isolator, the microwave is split into two outputs using the power splitter where one of them is fed to a variable phase shifter in order to output subsequent pulses with two relative phases, {\it e.g.} X and Y phases.
The phase is continuously controllable with the accuracy of $\leq$2.4 degrees in 215$-$240 GHz and $\leq$1.2 degrees in 107$-$120 GHz.
In pulsed EPR operation, cw output of the synthesizer is gated using the PIN switch where the timing of the gating is controlled by transistor-transistor logic (TTL) signals from the pulse generator (see Fig.~1).
Typical rise and fall times of the PIN switch are 12 ns, which makes it possible to produce as short as $\sim$20 ns pulse.
After the microwave is amplified by the preamplifier, it goes into a frequency multiplier.
The frequency multiplier is made of a pair of the frequency multiplier stages in which each stage consists of active and passive frequency multipliers (VDI), an isolator and phase trimmers.
The synthesizer frequency is multiplied by 24 and 12 times for the output in the range of 215$-$240 and 107$-$120 GHz, respectively.
An equal power of the microwave split by a power splitter is transmitted to each of the frequency multiplier stage.
The relative microwave phase between the two frequency multiplier stages is adjusted to be in-phase by the phase trimmers,
then the outputs of the two stages are combined by a high-frequency power combiner (VDI).
The output power of the source is 30$-$100 mW in 215$-$240 GHz (200$-$700 mW in 107$-$120 GHz) where the peak power is 100 mW at 230 GHz (700 mW at 115 GHz).
In addition, for PELDOR measurements, the source can be configured to the double-frequency mode.
As described in Fig.~2(b), each synthesizer chain consists of the directional coupler, the isolator, the PIN switch, the pre-amplifier and the power splitter,
and the microwave from the both chains is sequentially transmitted into the pair of the frequency multiplier stages.
In the same manner as the single-frequency mode, the double-frequency mode provides the output frequency range of 215$-$240 GHz and 107$-$120 GHz
and the peak output power of 100 mW and 700 mW at 230 GHz and 115 GHz, respectively.
Therefore, the source provides a high-power and extremely wide range of the tunable frequency for EPR and PELDOR spectroscopy.

\subsection{Quasioptical system}
A quasioptical system is employed to guide high-frequency excitations from the source to a sample and EPR signals to a mixer detector.
Figure~3 shows an overview of the quasioptical system composed of corrugated horns, wire-grid polarizers, Faraday rotators, corrugated waveguide (Thomas Keating), right-angle ellipsoid mirrors and transmitter and receiver stages (fabricated by the USC machine shop).
In order to cancel frequency dependence of the quasioptics, the quasioptical system is designed as a periodic focusing system with a 508 mm period using a Gaussian mode ray analysis~\cite{kogelik66, goldsmith}.
As shown in Fig.~3, the source outputs first couple to a corrugated horn in which single-mode outputs from a WR-3.4 (WR-8.0) rectangular waveguide of the source for 215$-$240 GHz (107$-$120 GHz) transition to the HE$_{11}$ mode, then efficiently couple to the fundamental Gaussian mode at the output of the horn (coupling efficiency between the HE$_{11}$ mode and the Gaussian mode is $\sim$99 $\%$)~\cite{goldsmith}.
Gaussian waves are guided by the right-angle ellipsoid mirrors with f=254 mm. The intensity of high-frequency excitation waves is controlled by a combination of a rotating and fixed-angle wire-grid polarizers.
We also employ a quasioptical isolator consisting of a combination of the fixed-angle wire-grid polarizer and the Faraday rotator to prevent standing waves in the transmitter stage.
After the second ellipsoid mirror, Gaussian waves go through the circulator (wire-grid), then couple to the HE$_{11}$ mode in the corrugated waveguide.
A sample is located at the bottom end of the waveguide.
The circulator based on the wire-grid provides linearly polarized excitations to the sample.
For detection of EPR signals, the system employs the induction-mode detection~\cite{smith98} where the circulator separates circularly polarized EPR signals from the linearly polarized excitations (typical isolation more than 30 dB).
Then EPR signals propagate the quasioptics in the receiver stage, and couple to the mixer detector through a corrugated horn.
The receiver stage also has a similar quasioptical isolator to reduce standing waves,
and is enclosed by an isolation box to isolate it from background noises, {\it e.g.} scattered excitation waves from the quasioptics.
The transmitter and receiver stages are separately mounted to optimize the coupling to the corrugated waveguide independently.

\subsection{Detection system}
For detection of EPR signals, we built a phase-sensitive superheterodyne detection system (see Fig.~4).
In the detection system, EPR signals from the quasioptical system are first down-converted to 3 GHz IF signals by a subharmonically pumped mixer (VDI) and the high-frequency LO (VDI).
The high-frequency LO consists of a microwave synthesizer (2$-$20 GHz, Micro Lambda Wireless), a directional coupler (ATM), a PIN switch (American Microwave), an isolator (Ditom Microwave) and frequency multipliers.
IF signals are immediately amplified by a low-noise amplifier (LNA, noise figure=0.5 dB, MITEQ), then by a second amplifier (AML Communications).
As shown in Fig.~4, a 3 GHz reference is produced by a reference circuit including a mixer (Marki Microwave), amplifiers, a $\times$24/$\times$12 frequency multiplier (Mini Circuits), a phase shifter and variable attenuator (ATM).
In the reference circuit, a 125/250 MHz reference synthesized by mixing Reference 1 from the source and Reference 2 from the LO is fed into the $\times$24/$\times$12 frequency multiplier to produce the 3 GHz reference signals for 230/115 GHz operation, respectively. Finally IF signals at 3 GHz are down-converted to I and Q d.c. signals by mixing with the 3 GHz reference with a IQ mixer (Marki Microwave). In cw EPR measurements, the power of IF signals is adjusted by a variable attenuator (GT Microwave) to optimize the mixer response, then intensities of I and Q signals are measured by lock-in amplifiers.
In pulsed EPR measurements, transient I and Q signals are sampled simultaneously by a fast digital oscilloscope.

\subsection{12.1 Tesla cryogenic-free superconducting magnet}
The HFEPR spectrometer employs a 12.1 Tesla cryogenic-free superconducting magnet (Cryogenic Limited).
The magnet consists of a sweepable 12 Tesla main coil and 0.1 Tesla sweep coil controlled by a separate magnet power supply.
The magnet is cooled by a single cryocooler (Cryomech), which operates at 2.8K with no load and is surrounded by a radiation shield at 40K.
The cryocooler is specified to operate in the stray field of the magnet and provide in excess of 20,000 hours continuous operation.
The magnet has a 89-mm room temperature bore for the access of a sample and a variable temperature control system.

\subsection{Sample holder and cryostat}
In the present HFEPR setup, we employ sample holders without a cavity to measure various forms of samples, {\it e.g.} single crystals, thin films,
powders and frozen solutions~\cite{Takahashi12}.
Single crystals and thin films are directly placed on a conductive end-plate and are positioned inside or near the end of the corrugated waveguide.
For powder or frozen solution samples, we use a "bucket" with a volume of 12$-$527 microliters to hold the samples, {\it e.g.} the dimensions of 12 (527) microliter buckets are 2.0 (9.4) mm of the diameter and 3.8 (7.6) mm of the height.
The bucket is made of Teflon which is placed on the end-plate.

Sample temperature is controlled by a $^4$He cryostat (Janis) inserted in the magnet bore. The cryostat has a 62 mm diameter for sample access and an optical window at the bottom for excitation and detection of visible and infrared light. Liquid Helium (LHe) is stored in a LHe reservoir integrated in the cryostat and the flow rate of LHe is adjusted by a rotary vane pump and a pressure regulator. By controlling the temperature of the flowing helium vapor in which samples are immersed, both the sample and holder are simultaneously cooled to a desired temperature, thereby eliminating the need for thermal anchoring and sample mount heating. The LHe flow rate and a heater current in the system are balanced to provide flowing vapor and sample temperatures over the range of 300-1.4 K. A temperature controller (Lake Shore) together with application of two temperature sensors at the cryostat and the sample provides precise and stable temperature control.

\section{RESULTS AND DISCUSSION}
In this section, we assess pulsed EPR sensitivity of the HFEPR spectrometer. We also present two examples of unique experiments which were recently performed using the HFEPR spectrometer.

\subsection{Pulsed EPR sensitivity}
Pulsed EPR sensitivity of the HFEPR spectrometer was estimated using spin echo measurement.
For the estimate, a thin film ($\sim$1.6$\times$1.6$\times$0.3 mm$^3$) of 1 wt.$\%$ 1:1 complex of a,c-bisdiphenylene-b-phenylallyl (BDPA) and benzene mixed in polystyrene was used.
BDPA is a stable radical ($S$=1/2 and $g$=2.003) and is widely used in EPR and related techniques.
Based on the measured weight (0.7$\pm$0.1 mg), the number of spins in the film was calculated to be 8$\times$10$^{15}$-1$\times$10$^{16}$.
As shown in Fig.~5(a), 115 GHz cw EPR spectrum of the BDPA film is represented by a single EPR signal with 0.9 mT peak-to-peak linewidth.
Figure~5(b) shows a single-shot spin echo signal measured at $T$=300 K and $B$=4.110 T with 2$\tau$=2.1 $\mu$s.
As shown in the inset of Fig.~5(b), $T_2$=0.62 $\mu$s of the BDPA film was extracted from the decay of the spin echo signals by varying 2$ \tau$.
Here we estimate pulsed EPR sensitivity using the spin echo signal in Fig.~5(b) and the following equation~\cite{Hertel05, Morley08},
\begin{equation}
Sensitivity=\frac{Nf}{SNR}e^{-2\tau/T_2}
\end{equation}
where $N$ is the number of spins, $f$ is the fraction of the spins that are excited, $SNR$ is the signal-to-noise ratio of spin echo signal and $\tau$ is the pulse separation in the spin echo measurement.  $SNR$ of the single-shot spin echo signal was 14. The fraction of the spins that are excited is given by the ratio of excitation bandwidth to EPR linewidth. The $\pi$ pulse of 300 ns corresponding to the excitation bandwidth of 0.12 mT and the linewidth of 0.9 mT gives $f$=0.1.
Thus, pulsed EPR sensitivity of the HFEPR spectrometer at 115 GHz was estimated to be 2-3$\times$$10^{12}$ spins at $T$=300 K where the polarization of BDPA spins is 0.9 $\%$.
This sensitivity corresponds to 2-3$\times$$10^{10}$ spins at $T$=2 K where the polarization is 88 $\%$.
Similarly, pulsed EPR sensitivity at 230 GHz was estimated to be 5-6$\times$$10^{11}$ and 9$\times$$10^{9}$-1$\times$$10^{10}$ spins at $T$=300 and 2 K where the spin polarization is 1.8 and 99 $\%$, respectively.

\subsection{PELDOR measurement}
Here we employ a PELDOR technique to detect couplings between a probe spin and surrounding spins.
The sample studied here was a commercially available (Sumitomo Electric Industries) high-temperature high-pressure (HTHP) type-Ib diamond crystal with two polished parallel (111) surfaces and nominal dimensions of 1.5$\times$1.5$\times$1.0 mm$^3$.
The diamond crystal contained substitutional single-nitrogen paramagnetic impurities (N) with concentration of 10$-$100 ppm, corresponding to average spatial separation of 4$-$8 nm among the impurities. Recent studies indicates that dipolar interactions between N spins are responsible to spin decoherence time $T_2$ in type-Ib diamond crystals~\cite{kennedy03, Takahashi08, Wang13}.
Figure~6(a) shows echo-detected field sweep spectrum of the diamond crystal at 230 GHz with application of static magnetic field along (111) axis of the diamond lattice.
As shown Fig.~6(a), the spectrum shows five pronounced peaks representing N centers in diamond ($S$=1/2, $I$=1, $A_{x,y}$=82 MHz and $A_z$=114 MHz)~\cite{Loubser78}.
The five peak EPR signal is originated that there are four bond axes of N centers, {\it i.e.} (111), ({\={1}}11), (1{\={1}}1) and (11{\={1}}) and
the magnetic field  was applied along (111) in the measurement~\cite{Takahashi08}.
With pulsed EPR measurements, we also found that spin decoherence time and spin-lattice relaxation time of the diamond crystal were $T_2$$\sim$1 $\mu$s and $T_1$$\sim$2 ms respectively.
Next we applied a three-pulse PELDOR sequence to probe dipolar couplings between N spins in diamond~\cite{Milov81}.
As shown in the inset of Fig.~6(b), the applied PELDOR sequence consists of the spin echo sequence for probe spins and a single $\pi$ pulse for pump spins.
In PELDOR spectroscopy, changes in spin echo signal occurs when effective magnetic dipolar field at probe spins is altered by flipping pump spins with a $\pi$ pulse.
In the measurement, we chose N spins on resonance at $B$=8.211 T as probe spins ({\it i.e.} N centers whose bonds are along (111) and whose state is $|m_I=-1>$;
the spin state is conserved during the measurement because $T_1$$\gg$$T_2$), and other N spins as pump spins.
As shown in Fig.~6(b), PELDOR signals were observed clearly in the echo intensity of probe spins as a function of pump pulse frequency.
The number of PELDOR signals and the splitting of the peaks correspond to N spins,
which indicate a direct observation of the dipolar coupling between N spins in the type-Ib diamond crystal.

\subsection{Application of DD techniques}
Coherence of spin systems can be extended by reducing couplings between the spins and surrounding noises using advanced pulse sequences such as DD sequences.
The idea of DD can be traced into the spin echo, in which couplings to static and spatially inhomogeneous magnetic field noises can be suppressed completely by the rephasing pulse in the middle of the elapsing time interval~\cite{hahn50}.
The efficacy of DD depends on the relationship between the noise spectrum of surrounding spins and the spectrum of a DD sequence.
When those two spectra overlap largely, spins experience strong decoherence~\cite{Bylender11, Wang13}.
Here we describe the demonstration of DD sequences to N spins in diamond at 115 GHz.
The sample we investigated here was another HTHP type-Ib diamond (Element Six) with two polished parallel (100) surfaces and nominal dimensions of 0.8$\times$0.8$\times$0.4 mm$^3$.
The concentration of N impurities in the diamond sample was 10$-$100 ppm.
Spin decoherence time $T_2$ of the diamond sample was first determined to be 23$\pm$1 $\mu$s by fitting a spin echo decay with a single exponential function.
We investigated the application of DD sequences including Carr-Purcell-Meiboom-Gill (CPMG) sequence~\cite{Carr54, Meiboom58}, two-axis CPMG sequence (denoted as XY) and Uhrig dynamical decoupling (UDD) sequence~\cite{Uhrig07}.
As shown in Fig.~7(a), excitation pulses of the sequences consist of one $\pi$/2 and four $\pi$ pulses (denoted as $N$=4 where $N$ is the number of $\pi$ pulses).
All of $\pi$ pulses in CPMG and UDD sequences have Y phase while the phase of $\pi$ pulses in XY sequence alternates between X and Y.
Coherence time ($T_{coh}$) of the diamond sample was determined by fitting the echo decay over the free evolution time (Fig.~7(b)) with a single exponential function.
As indicated by comparison of spin echo decay with CPMG, XY and UDD for $N$=8 (see Fig.~7(b)), we found a longer coherence time ($T_{coh}$) with CPMG and XY than that with UDD.
Finally Figure~7(c) shows spin echo decays as a function of the free evolution time obtained with CPMG sequences with $N$ $\pi$ pulses. We found that application of 128 $\pi$ pulses extends $T_{coh}$ to 211$\pm$8 $\mu s$, corresponding to 9 fold enhancement compared to spin decoherence time $T_2$ measured by spin echo sequence.

\section{ACKNOWLEDGMENTS}
We would like to thank Chathuranga Abeywardana, Steven Retzloff, Jeffrey Hesler, Stephen Jones and Don Wiggins for technical
assistance. This work was supported by the Searle scholars program (S.T.).


\clearpage

\noindent{{\bf Figure captions}}

\bigskip

Fig.~1.
Schematic overview of the HFEPR spectrometer at USC. A quasioptical system consisting of transmitter and receiver stages and a corrugated waveguide is employed to guide excitations and EPR signals. EPR signals are detected by a phase-sensitive detection system. Reference for the detection system is produced from the source (Ref 1) and the LO (Ref 2). Timing of pulsed excitations and data acquisition by a digital oscilloscope is controlled by a pulse generator.

Fig.~2.
Schematic overviews of the high-frequency high-power solid-state source in the HFEPR spectrometer. (a) Configuration of single-frequency mode used for cw EPR, spin echo and DD measurements. (b) Configuration of double-frequency mode used for PELDOR measurements.

Fig.~3.
Overview of the quasioptical system consisting of transmitter and receiver stages. The quasioptical system is a periodic focusing system with the focal length of 254 mm for Gaussian waves. HE$_{11}$ mode in a corrugated horn connected to the source excites the Gaussian mode with high efficiency (coupling between HE$_{11}$ mode and the Gaussian mode is $\sim$99 $\%$). The periodic focusing system also allows using the same quasioptics for a wide range of frequencies. The isolation box to reduce background noises is made of aluminum and the inside is covered by absorbers.

Fig.~4.
Circuit diagram of the detection system. A subharmonically pumped mixer for EPR detection provides an excellent isolation of the LO to the RF port (\textgreater 100 dB). The high-frequency LO consists of a broad-band microwave synthesizer (2-20 GHz, Micro Lambda Wireless), a directional coupler, an isolator, a PIN switch, an amplifier and a frequency multiplier. The PIN switch in the LO is for protection of the detection system. The superheterodyne detection has $\sim$1 GHz bandwidth and its noise temperature ($T_N$) is $\sim$1200 K. IF power is controlled by a variable attenuator to optimize the power to the IQ mixer.

Fig.~5.
EPR measurements of 1 wt.$\%$ BDPA in polystyrene at room temperature. (a) cw EPR spectrum of the BDPA film showing a single EPR signal with 0.9 mT peak-to-peak linewidth. The spectrum was taken by single scan at 0.1 mT/s with 20 kHz-0.1 mT of the field modulation. (b) A single-shot spin echo signal with 2$\tau$=2.1 $\mu$s. A $\pi$/2 pulse of 200 ns and a $\pi$ pulse of 300 ns were used. The excitation pulses is largely truncated in the plot. The inset shows spin echo intensity as a function of 2$\tau$. $T_2$=0.62 $\mu$s was obtained by fitting the decay with a single exponential function.

Fig.~6.
(a) Echo-detected field sweep spectrum of the type-Ib diamond crystal at 230 GHz.
EPR signals of N spins are shown as five peaks.
(b) PELDOR measurement in the type-Ib diamond.
The magnetic field ($B$) was fixed at 8.211 T and echo intensity was measured while the frequency of a $\pi$ pulse for pump spins was swept. The inset shows the applied pulse sequence.
A $\pi$/2 pulse of 300 ns and a $\pi$ pulse of 500 ns were used in the measurement.
$\pi$ pulses are represented by solid squares and $\pi$/2 pulses are represented by open squares.

Fig.~7.
(a) DD sequences with $N$=4 where $N$ is the number of $\pi$ pulses. Top: Carr-Purcell-Meiboom-Gill (CPMG) sequence, middle: two-axis CPMG sequence (denoted as XY) and bottom: Uhrig dynamical decoupling (UDD) sequence. $\pi$ pulses are represented by solid squares and $\pi$/2 pulses are represented by open squares. Phases of excitation pulses are set with respect to the phase of the reference microwave in the detection system.
(b) Application of CPMG, XY and UDD for $N$=8 at 115 GHz. Data with errors are represented by markers with designated shape as shown in the legend and fits to a single exponential function are shown by solid lines with designated color. $T_{coh}$ was measured to be 65$\pm$1, 67$\pm$3 and 35$\pm$2 $\mu$s with CPMG, XY and UDD, respectively. The inset shows a trace of spin echo signals with CPMG for $N$=4. The echoes are represented with solid lines.
(c) Dependence of $T_{coh}$ on $N$ with CPMG sequences at 115 GHz. Data with errors are represented by markers with designated shape as shown in the legend and fits to a single exponential function are shown by solid lines with designated color. $T_{coh}$ was measured to be 23$\pm$1, 42.1$\pm$0.5, 48.8$\pm$0.8, 65$\pm$1, 90$\pm$1, 118$\pm$2, 168$\pm$3 and 211$\pm$8 $\mu$s for $N$=1 (spin echo), 2, 4, 8, 16, 32, 64 and 128, respectively.

\clearpage
{\bf Figure 1:}
\begin{figure}
\includegraphics[width=100 mm]{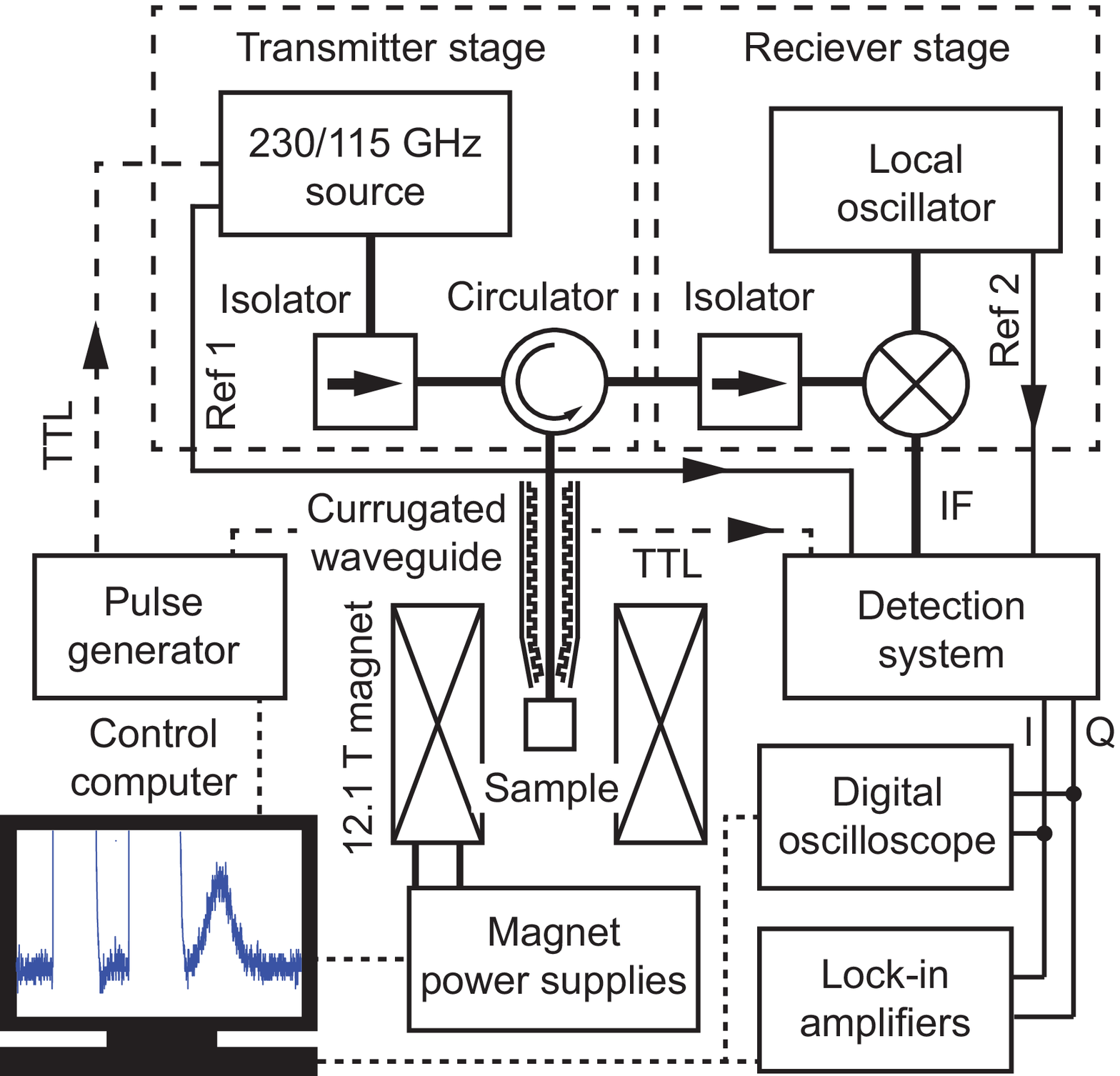}
\end{figure}

\clearpage
{\bf Figure 2:}
\begin{figure}
\includegraphics[width=150 mm]{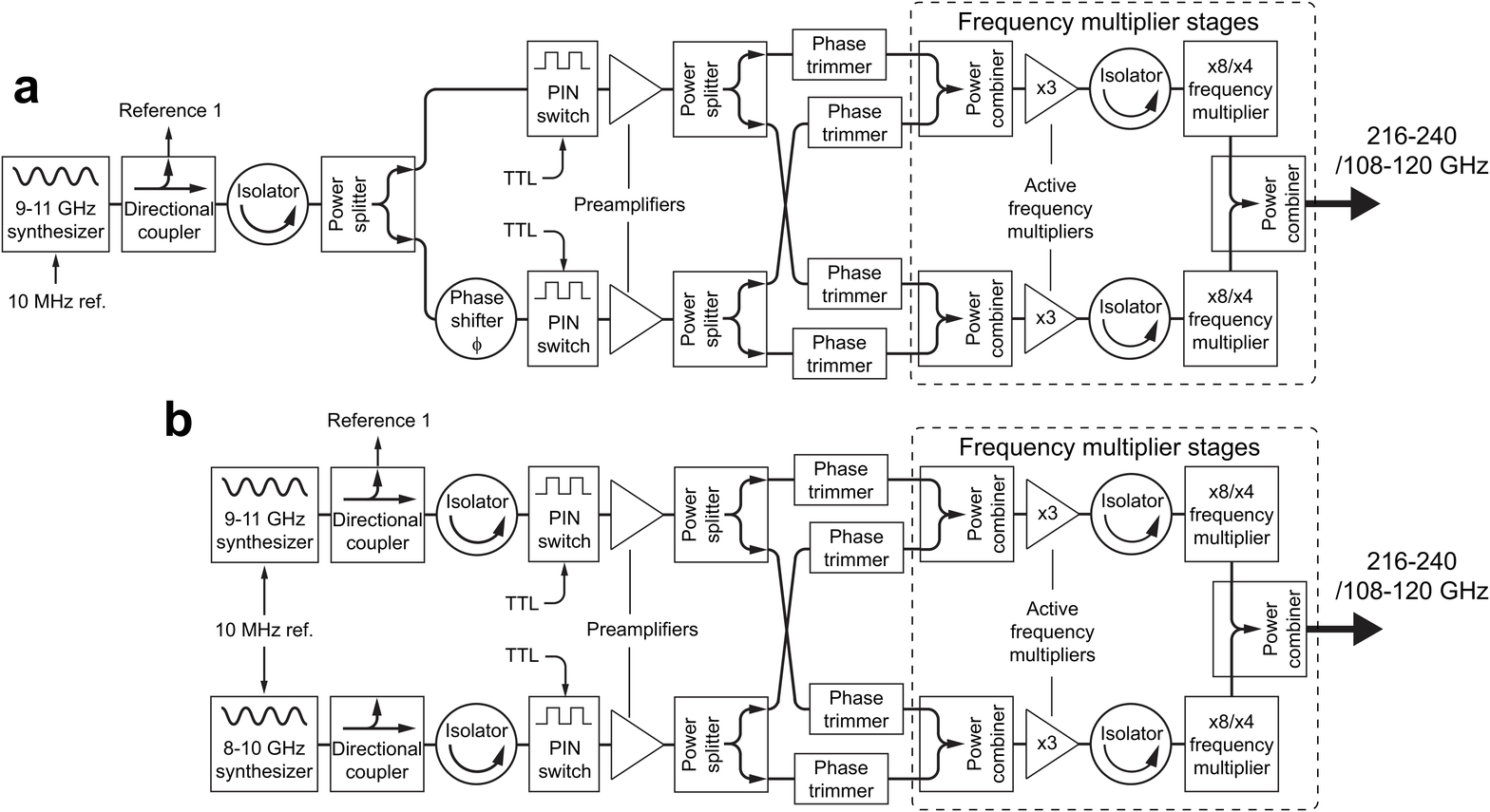}
\end{figure}

\clearpage
{\bf Figure 3:}
\begin{figure}
\includegraphics[width=150 mm]{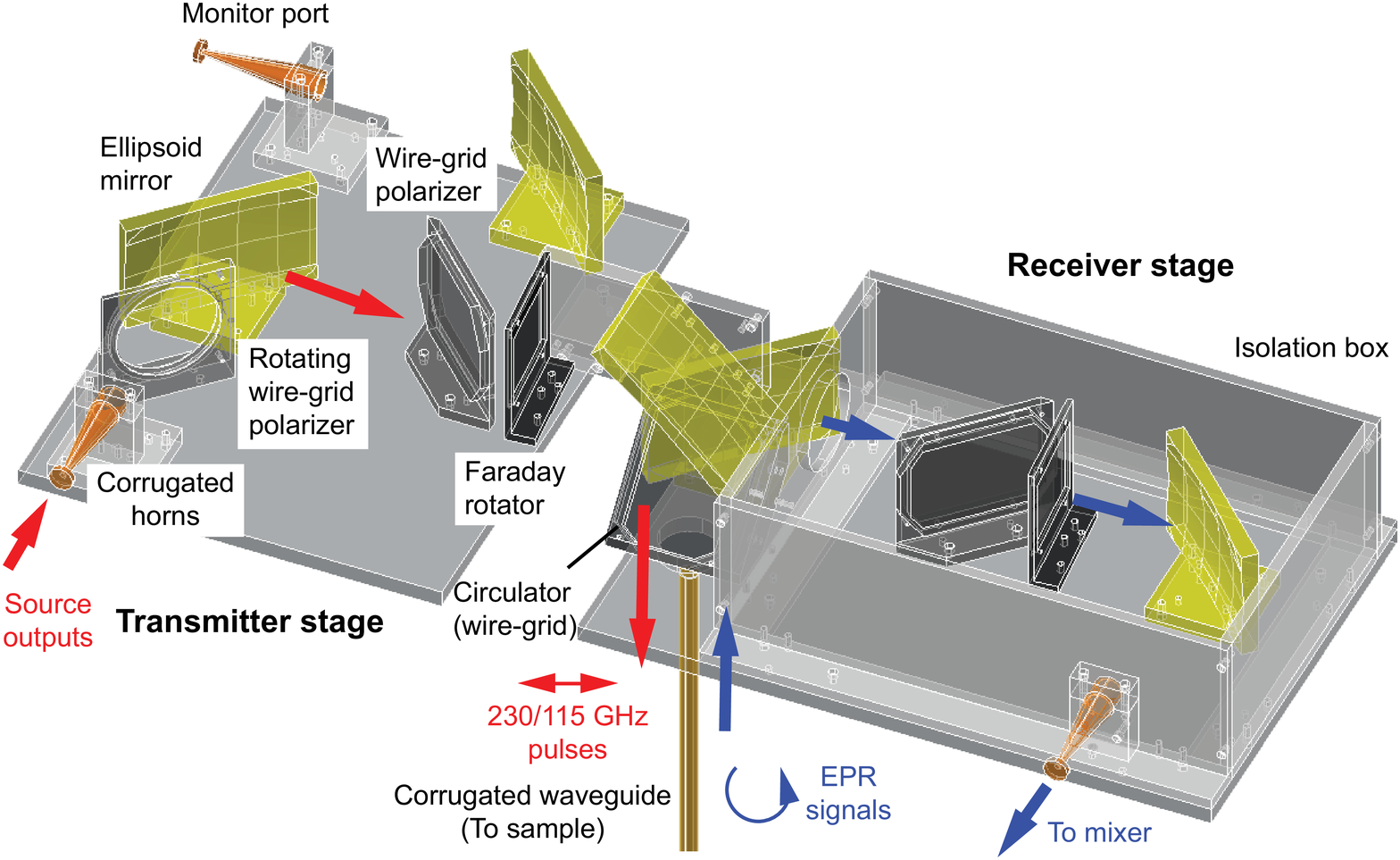}
\end{figure}

\clearpage
{\bf Figure 4:}
\begin{figure}
\includegraphics[width=150 mm]{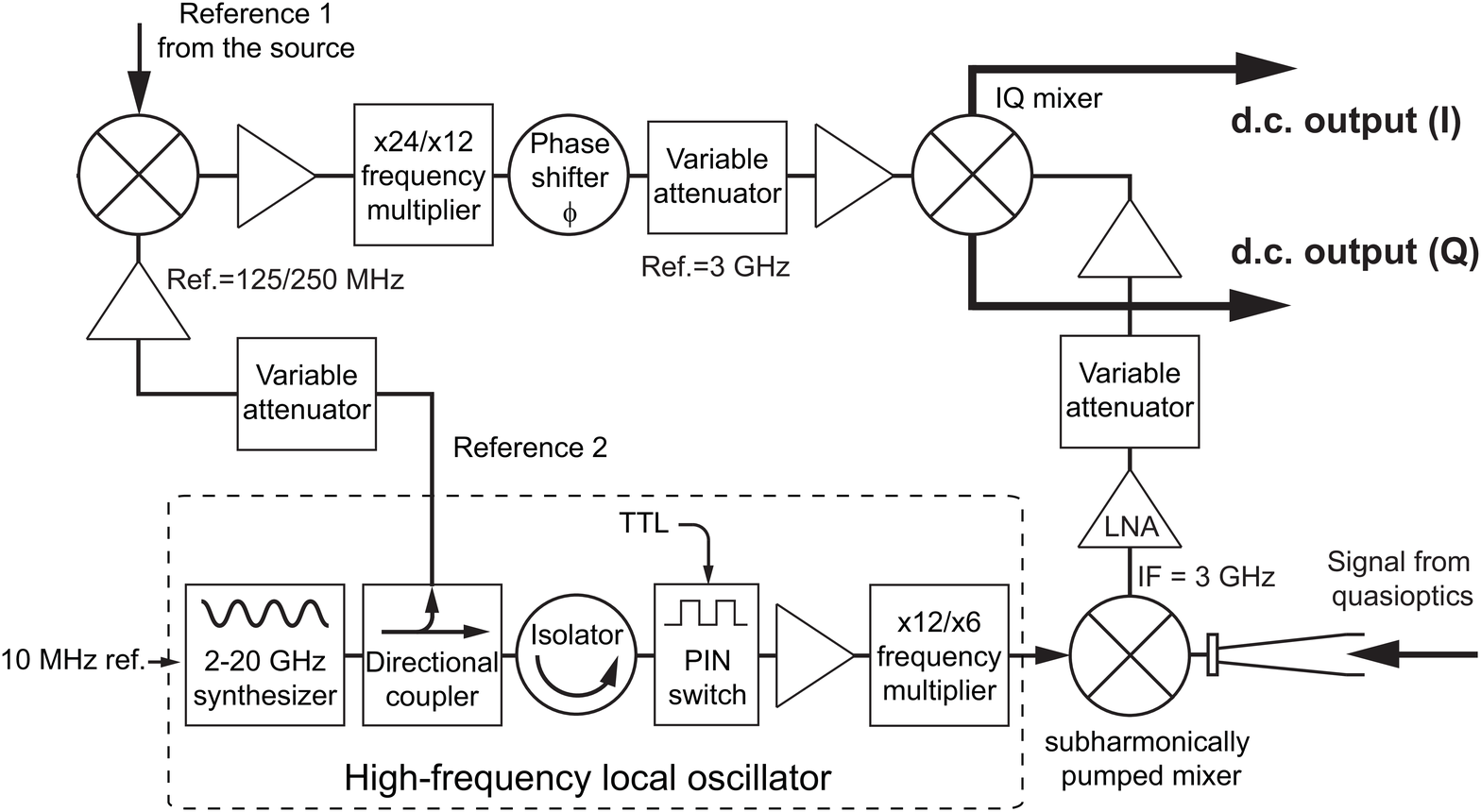}
\end{figure}

\clearpage
{\bf Figure 5:}
\begin{figure}
\includegraphics[width=90 mm]{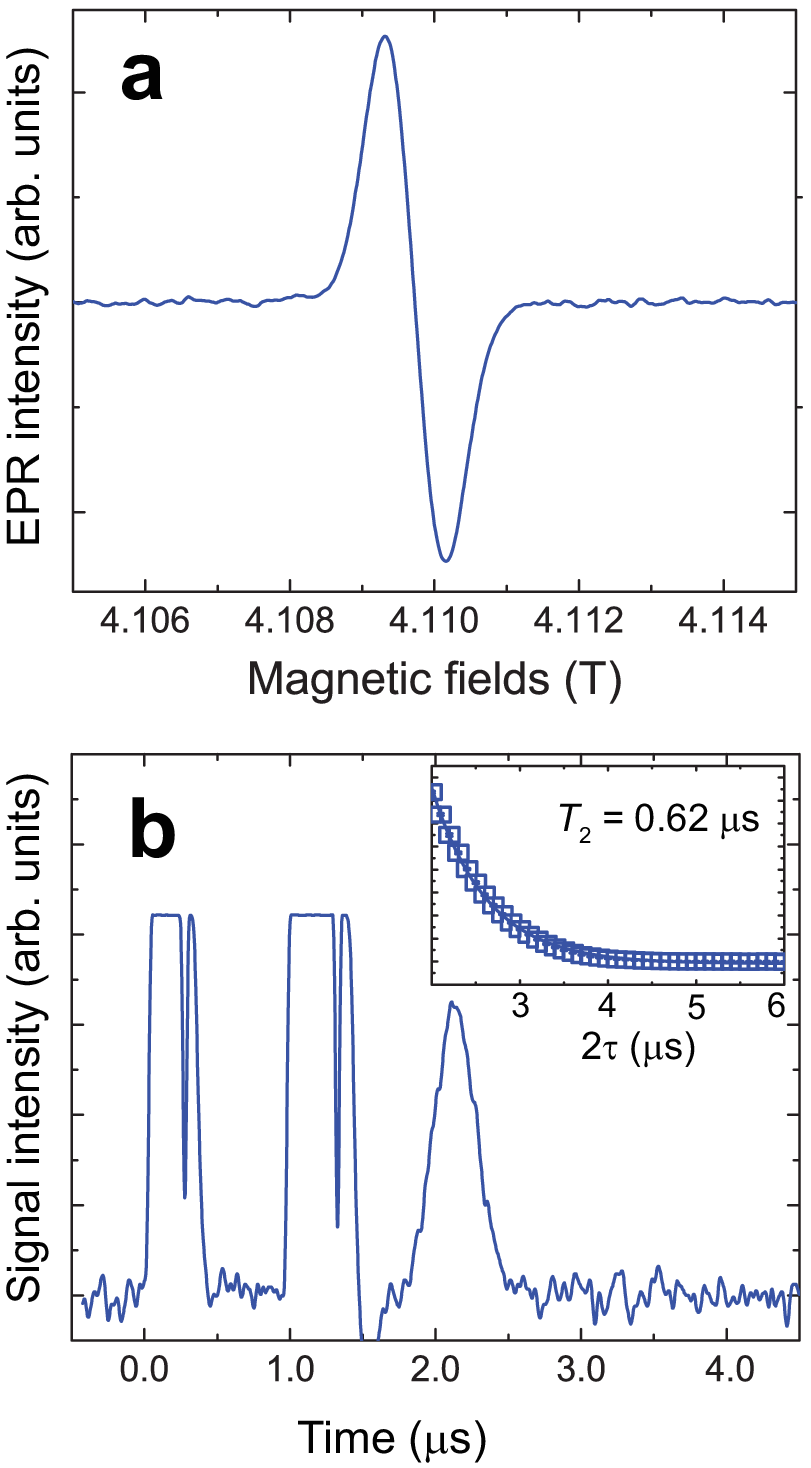}
\end{figure}

\clearpage
{\bf Figure 6:}
\begin{figure}
\includegraphics[width=90 mm]{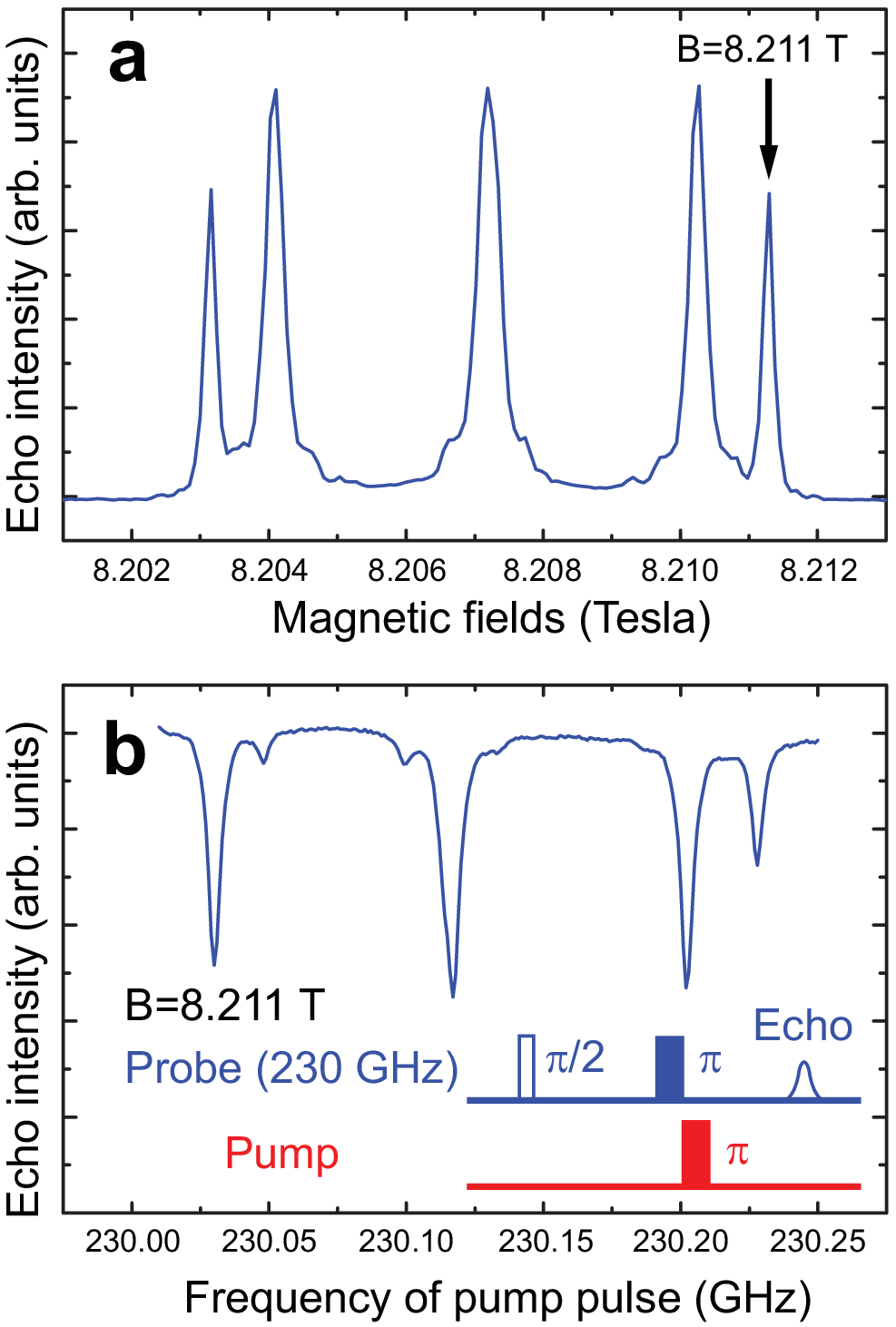}
\end{figure}

\clearpage
{\bf Figure 7:}
\begin{figure}
\includegraphics[width=90 mm]{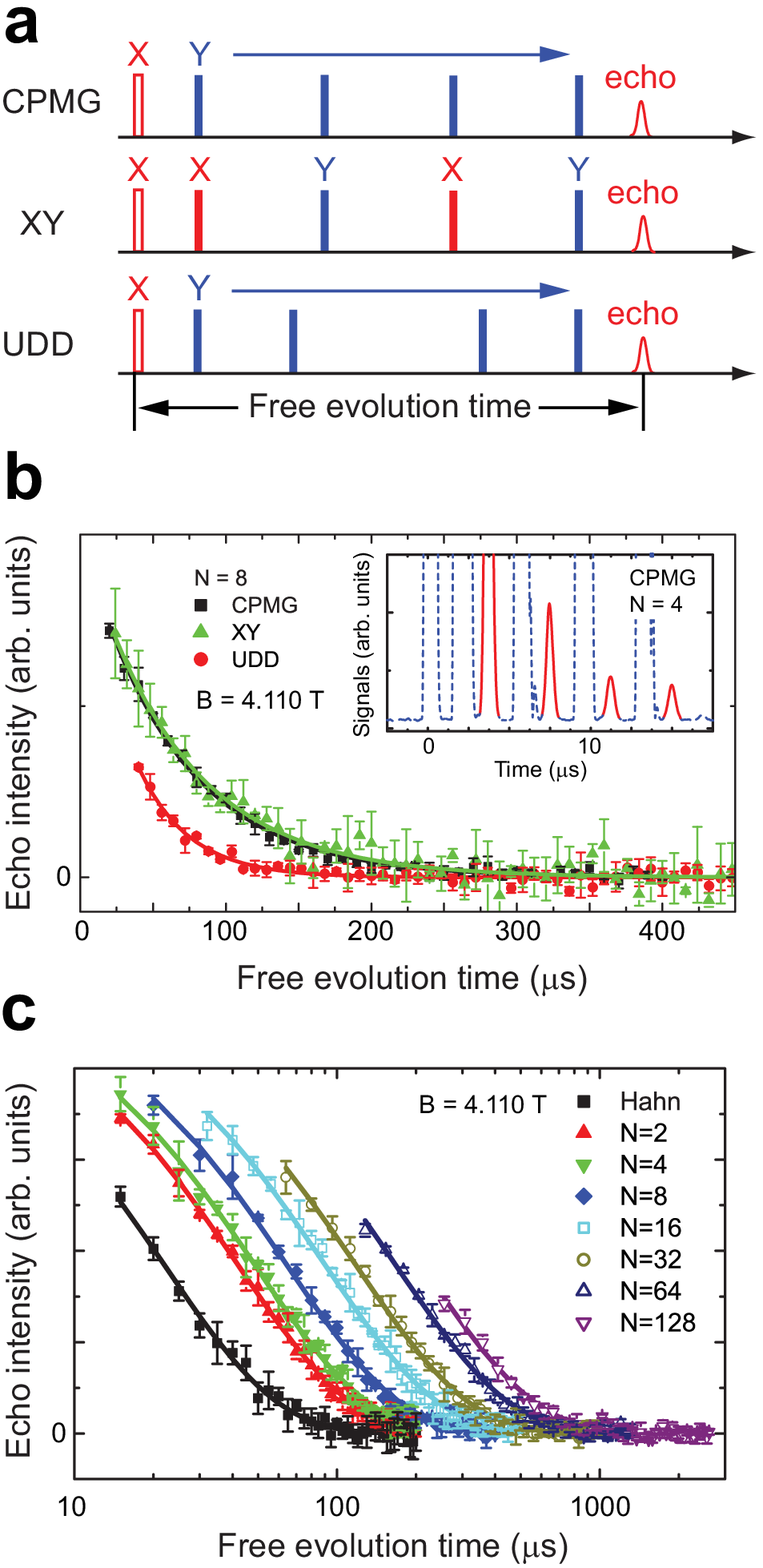}
\end{figure}

\end{document}